\documentclass[12pt,preprint]{aastex}

\slugcomment{To be published in ApJ}

\shorttitle{The CO Outflows of IRAS 16293--2422}

\shortauthors{Yeh et al.}

\begin{document}

\title{The CO Molecular Outflows of IRAS 16293--2422 Probed by the
Submillimeter Array}

\author{Sherry C. C. Yeh\altaffilmark{1,2}, 
Naomi Hirano\altaffilmark{1},
Tyler L. Bourke\altaffilmark{3},
Paul T. P. Ho\altaffilmark{1,3}
Chin-Fei Lee\altaffilmark{1,4},
Nagayoshi Ohashi\altaffilmark{1},
Shigehisa Takakuwa\altaffilmark{1,5}
}

\altaffiltext{1}{Institute of Astronomy \& Astrophysics, Academia Sinica,
P.O.  Box 23-141, Taipei 106, Taiwan, R.O.C.} 

\altaffiltext{2}{Current Address: Department of Astronomy \& Astrophysics,
University of Toronto, 50 St. George Street, Toronto, ON M5S 3H4, Canada
; yeh@astro.utoronto.ca}

\altaffiltext{3}{Harvard-Smithsonian Center for Astrophysics, 60 Garden
Street, MS 78, Cambridge, MA 02138, USA}

\altaffiltext{4}{Harvard-Smithsonian Center for Astrophysics, 
Submillimeter Array, 645 North A'ohoku Drive, Hilo HI 96720, USA}

\altaffiltext{5}{National Astronomical Observatory of Japan, ALMA Project
Office, Osawa 2-21-1, Mitaka, Tokyo, 181-8588, Japan}

%\doublespace
\begin{abstract}

We have mapped the proto-binary source IRAS 16293--2422 in CO 2--1,
$^{13}$CO 2--1, and CO 3--2 with the Submillimeter Array (SMA).  The maps
with resolution of 1\farcs5--5\arcsec\ reveal a single small scale
($\sim$3000 AU) bipolar molecular outflow along the east-west direction.
We found that the blueshifted emission of this small scale outflow mainly
extends to the east and the redshifted emission to the west from the
position of IRAS 16293A. 
%This suggests that the small scale outflow detected by the SMA is most
%likely driven by IRAS 16293A.  
A comparison with the morphology of the large scale outflows previously
observed by single-dish telescopes at millimeter wavelengths suggests that
the small scale outflow may be the inner part of the large scale
($\sim$15000 AU) E--W outflow.  On the other hand, there is no clear
counterpart of the large scale NE--SW outflow in our SMA maps.  Comparing
analytical models to the data suggests that the morphology and kinematics
of the small scale outflow can be explained by a wide-angle wind with an
inclination angle of $\sim$30\degr--40\degr~with respect to the plane of
the sky.  The high resolution CO maps show that there are two compact,
bright spots in the blueshifted velocity range.  An LVG analysis shows that
the one located 1\arcsec\ to the east of source A is extremely dense,
n(H$_2$)~$\sim$10$^{7}$ cm$^{-3}$, and warm, T$_{\rm kin} >$~55 K.  The
other one located 1\arcsec\ southeast of source B has a higher temperature
of T$_{\rm kin} >$~65 K but slightly lower density of
n(H$_2$)~$\sim$10$^{6}$ cm$^{-3}$.  It is likely that these bright spots
are associated with the hot core-like emission observed toward IRAS 16293.
Since both two bright spots are blueshifted from the systemic velocity and
are  offset from the protostellar positions, they are likely formed by
shocks.

\end{abstract}


\section{Introduction}

Outflow phenomena have been recognized as an important phase in the star
formation processes.  It is generally accepted that when stars are formed
by gravitational infall, outflows transfer excess angular momentum out of
the system  \citep[e.g.,][]{shu1987}.  Outflows are observed at various
wavelengths from the radio to the ultraviolet, among which the
millimeter/submillimeter bands probe the molecular outflows.  Molecular
outflows, recognized as high-velocity wings in CO and other
molecular lines, are considered to be the ambient gas entrained or pushed
by protostellar winds.  By observing molecular outflows, we can determine
their physical, kinematic and chemical properties, trace the mass-loss
history of protostars, and probe the early phases of the star formation
process (\citealt{arce2007}, and references therein).

IRAS 16293--2422 (hereafter referred to as I16293) is a well-studied
proto-binary system located in the nearby molecular cloud complex L 1689.
The distance to L 1689 is often assumed to be 160 pc \citep{whittet1974},
however some recent studies suggested a distance of $\sim$120 pc
\citep{degeus1989, knude1998}.  In this paper we adopt 120 pc as the
distance.  The projected separation of the two protostars is
$\sim$5\arcsec~\citep{mundy1992}, which then corresponds to 600 AU.  I16293
is classified as a Class 0 young stellar object (YSO) since this system is
undetectable at wavelengths shorter than 10 $\mu$m and shows a spectral
energy distribution (SED) with very low temperature \citep{andre1994}.  The
bolometric luminosity of I16293 is estimated to be $\sim$36 $L_\odot$
\citep{correia2004}.  I16293 is known to have a quadrupolar outflow
\citep{walker1988, mizuno1990}; one bipolar pair extends along the east
(blue) and west (red) (hereafter the E--W pair) directions, while the other
one has an orientation from northeast (red) to southwest (blue) (hereafter
the NE--SW pair) with an inclination angle of 30$^\circ$--45$^\circ$
\citep{hirano2001} or 65$^\circ$ \citep{stark2004}.  The NE--SW pair is
more collimated, and its axis crosses the position close to the SE
continuum source IRAS 16293A (hereafter source A).  Therefore this pair has
been thought to be powered by source A \citep[e.g.,][]{walker1988,
mundy1990}.  On the other hand, the E--W pair is not as well collimated and
its axis, though not well defined, seems to pass to the north of source A,
where the continuum source IRAS 16293B (hereafter source B) is the only
possible driving YSO. Hence source B has often been assumed as the driving
source of the E--W pair.  \citet{stark2004} showed that source B has very
narrow line widths, a low luminosity, and is not evidently associated with
high-velocity gas. They interpreted source B as a T--Tauri star that drove
the E--W pair and that the E--W pair is now a fossil flow.  However, these
interpretations are based on observations at low angular resolution (i.e.,
$>$5\arcsec), that are insufficient to resolve the two protostars and
reveal the relation between the quadrupolar outflow and the binary.

In fact, using higher angular resolution observations \citet{chandler2005}
proposed an alternative interpretation.  They resolved source A into four
components, two at 1~mm and another two at 7~mm, and suggested that two of
them are responsible for driving the NE--SW and the E--W pair,
respectively. In addition, \citet{chandler2005} proposed that source B is
actually much younger than previously interpreted and may have not yet
begun the phase of mass loss.  Their proposition was, however, made based
on (1) the dust continuum morphology and proper motions of source A, and
(2) the velocity structures of emission of H$_2$CO, H$_2$S, and SO, rather
than CO, which is a much better tracer of the overall outflow structure
free from chemical peculiarities and shocks. 

In order to study the molecular outflows in the vicinity of I16293 in more
detail, we have carried out CO 2--1, $^{13}$CO 2--1, and CO 3--2
observations at higher angular resolutions using the Submillimeter
Array\footnote{The Submillimeter Array is a joint project between the
Smithsonian Astrophysical Observatory and the Academia Sinica Institute of
Astronomy and Astrophysics and is funded by the Smithsonian Institution and
the Academia Sinica.} (SMA).  Our SMA observations provide sufficient
angular and spectral resolution in the CO lines to resolve the kinematics
and the structure of the outflow in the central region of I16293.  In this
paper, we summarize the observational details in \S2; present our SMA CO
2--1, $^{13}$CO 2--1, and CO 3--2  results in \S3; discuss the outflow
driving mechanism, physical parameters of the outflow, and the 
kinematics in the vicinity of I16293 in \S4. 

\section{Observations}

The Submillimeter Array (SMA) \citep{ho2004} observations of the CO J=2--1
(230.538 GHz), CO 3--2 (345.796 GHz), and $^{13}$CO 2--1 (220.399 GHz)
transitions were carried out during February 2003 to February 2005.  The
phase tracking center was $\alpha$(J2000)~=~16$^h$32$^m$22.91$^s$,
$\delta$(J2000)~=~$-$24$^{\circ}$28$'$35.52\arcsec.  The bandwidth of the
spectral correlator was 656 MHz in 2003 and 2 GHz in 2005.  The frequency
resolution used was 812.5 kHz, corresponding to the velocity resolution of 1.06
km s$^{-1}$ at 230 GHz and 0.70 km s$^{-1}$ at 345 GHz.  The field of view
of the array is $\sim$54\arcsec~at 230 GHz and $\sim$36\arcsec~at 345 GHz.
The CO 2--1 and $^{13}$CO 2--1 lines were observed simultaneously on 18
February 2005 with 6 antennas, providing 15 baselines ranging from 10 to 70
m.  The CO 3--2 data were observed on February 26 and April 23 in 2003 with
4 and 5 antennas, respectively, providing 10 independent baselines
ranging from 12 to 120 m.  We note that the CO 3--2 observations provide
fewer short baselines than the CO 2--1 observations, therefore the CO
3--2 data are less sensitive to extended structures.

%We note that we
%had less shorter baselines in the CO 3--2 observations than the CO 2--1
%observations, which makes the CO 3--2 observations less sensitive to
%extended structures as compared to the CO 2--1 observations. 

The visibility data were calibrated using the MIR software package
originally developed for Owens Valley Radio Observatory
\citep{scoville1993, qi2005} and modified for the SMA.  The absolute flux
calibration and bandpass calibration were done by observing Callisto (on
2003 February 26), Ganymede (on 2003 April 23), and Uranus (on 2005
February 18).  The uncertainty in the absolute flux scale is estimated to
be $\sim$10\%.  The visibility phase and amplitude calibration was achieved
by observing the quasars NRAO 530 (J1733-130) in 2003 and J1743--038 in
2005.  The continuum level was measured from line-free channels in the
calibrated visibility data.  The MIRIAD and the NRAO AIPS packages were
used to make the spectral images and for further analysis.  The maps were
restored using uniform weighting, resulting in synthesized beam sizes of
4\farcs7$\times$2\farcs5 with a position angle of 0.0$^{\circ}$ for the CO
2--1, 4\farcs9$\times$2\farcs6 with a position angle of 1.5$^{\circ}$ for
the $^{13}$CO 2--1, and 1\farcs6$\times$1\farcs4 with a position angle of
26.2$^{\circ}$ for the CO 3--2.  The rms noise levels of the channel maps
with $\sim$1 km s$^{-1}$ width are $\sim$0.15 Jy beam$^{-1}$ for the CO
2--1, $\sim$0.13 Jy beam$^{-1}$ for the $^{13}$CO 2--1, and $\sim$0.3 Jy
beam$^{-1}$ for the CO 3--2.

\section{Results}

\subsection{CO 2--1}

\begin{figure*}
%\epsscale{1.0}
\plotone{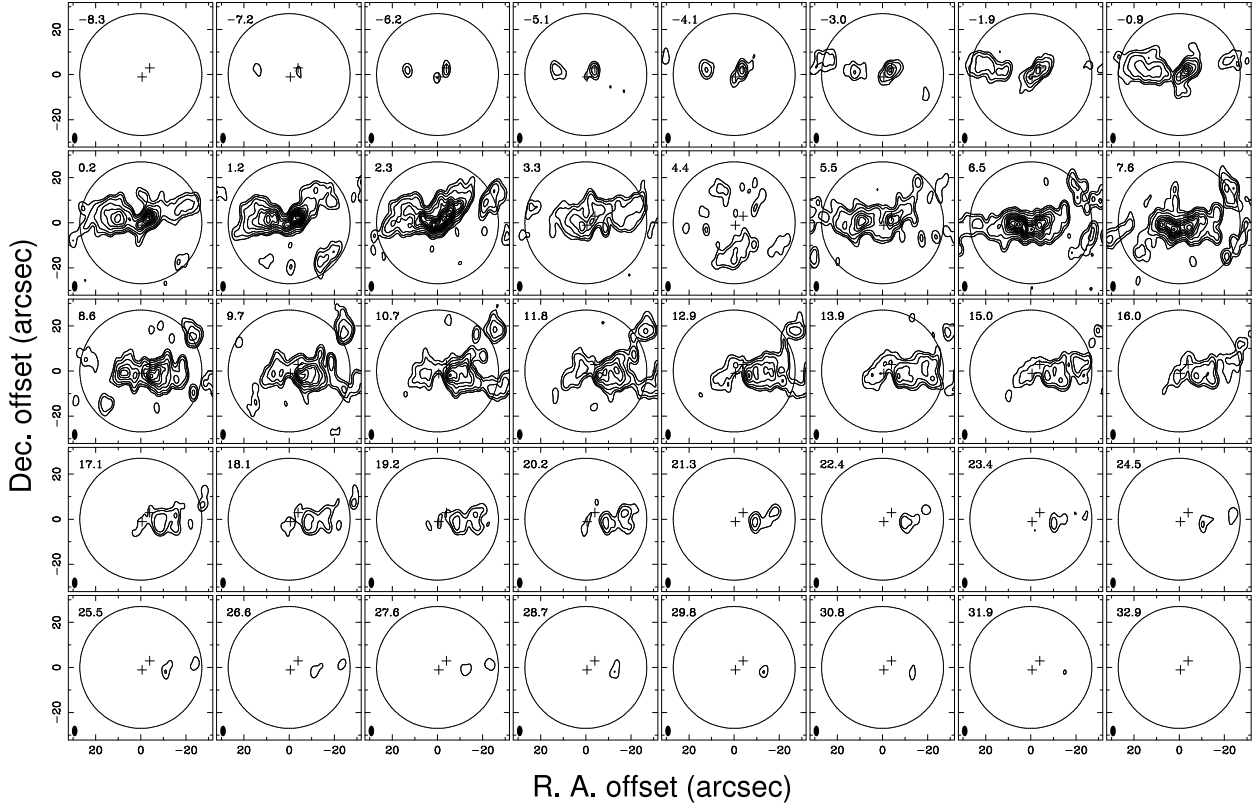}
\caption{Velocity channel maps of the CO 2--1 emission.  Offsets (in
arcseconds) are with respect to the phase tracking center,
$\alpha=$~16$^h$32$^m$22.91$^s$ and
$\delta=$~$-$24$^{\circ}$28$'$35.52\arcsec (J2000).  The contour levels
correspond to $-$5$\sigma$, 5$\sigma$, 10$\sigma$, 20$\sigma$, then
increase in every 20$\sigma$ step.  The 1$\sigma$ level is 0.15 Jy
beam$^{-1}$.  The crosses indicate the positions of the continuum sources A
and B; the open circle is the SMA field of view at 230 GHz, and the filled
ellipse indicates the synthesized beam.
Note that the primary beam correction is not applied to the data.
}
\label{f1}
\end{figure*}

Fig.~\ref{f1} shows the velocity channel maps of the CO 2--1 emission
uncorrected for the primary beam attenuation.  The
CO 2--1 emission is shown in the channels from $V_{\rm LSR}$ = $-$7.2 km
s$^{-1}$ to 31.9 km s$^{-1}$.  In most of the channels, the spatial
distribution of the CO 2--1 emission is elongated along the east-west
direction.  It is likely that this elongated CO emission originates from
the molecular outflow previously imaged on larger scales \citep{stark2004}.
The CO emission at $V_{\rm LSR}$ = 4.4 km s$^{-1}$ is scattered over the
field of view.  Since this velocity channel is close to the cloud systemic
velocity of $V_{\rm LSR}$ = 4.0 km s$^{-1}$ \citep{mizuno1990}, most of the
emission has been resolved out due to the lack of short baselines.  The CO
distribution at $V_{\rm LSR}$ = 3.3 km s$^{-1}$ and $V_{\rm LSR}$ = 5.5 km
s$^{-1}$ shows the east-west elongation, suggesting that the CO emission at
these channels comes from the outflowing gas even though the velocity
offset from the systemic velocity is small.  Therefore, we infer that the
CO 2--1 emission seen in the channels from $V_{\rm LSR}$ = $-$7.2 km
s$^{-1}$ to 3.3 km s$^{-1}$ is the blueshifted outflow component and that
in the channels from $V_{\rm LSR}$ = 5.5 km s$^{-1}$ to 31.9 km s$^{-1}$ is
the redshifted outflow component.

The blueshifted emission is extended to the east of source A and the west
of source B as well.  The eastern component is seen in most of the
blueshifted channels from $V_{\rm LSR}$ = $-$7.2 km s$^{-1}$ to 3.3 km
s$^{-1}$, while the western component is in the velocity range from $V_{\rm
LSR}$ = $-$0.9 km s$^{-1}$ to 3.3 km s$^{-1}$, and is closer to the
systemic velocity compared to the eastern component.  As shown in the
velocity channels from $V_{\rm LSR}$ = $-$0.9 km s$^{-1}$ to 1.2 km
s$^{-1}$, the eastern blueshifted component has a fan shape pointing to
source A with a position angle of $\sim$75$^{\circ}$. 

The redshifted emission extends to the east and the west of source A.  The
eastern redshifted component mostly overlaps with the eastern blue
emission.  
%In Fig.~\ref{f2},  the eastern redshifted component appears in the
%velocity range $V_{\rm LSR}=$~5.5 km s$^{-1}$ to 11.8 km s$^{-1}$.  
The western redshifted component has a fan shape structure with its apex at
the position of source A at P.A. $\sim$90$^\circ$.  This structure appears
from $V_{\rm LSR}$ = 8.6 km s$^{-1}$, and the emission moves away from the
source as the velocity increases.  
%Note that the primary beam correction is
%not applied to the images displayed in Fig.~\ref{f1} and Fig.~\ref{f2}.

The total integrated intensity (zeroth moment) maps of the blueshifted and
redshifted CO 2--1 line emission are presented in Fig.~\ref{f2}a
(uncorrected for primary beam attenuation).  In the blueshifted component,
there are two prominent peaks 1\arcsec~to the east of source A (labeled b1)
and 1\arcsec~southeast of source B (labeled b2).  The peak brightness
temperature of CO 2--1 emission at b1 and b2 are 57 K and 63 K,
respectively.  These two peaks are seen in all the blueshifted channels
(Fig.~\ref{f1}) starting from $V_{\rm LSR}$ = $-$6.2 km~s$^{-1}$.  We note
that both b1 and b2 have not been resolved in the previous CO outflow maps
observed with the single-dish telescopes \citep{stark2004}.  

In Fig.~\ref{f3}a we compare the CO 2--1 spectrum observed by the SMA with
that observed by the JCMT.  We have applied the primary beam correction,
smoothed the SMA map to have a 21\arcsec~Gaussian beam,  which is same as
the JCMT beam at 230 GHz, and obtained the spectrum at the phase tracking
center.  The JCMT spectrum was resampled to have the same velocity
resolution as the SMA data.  To convert the antenna temperature $T_{\rm
A}$* into brightness temperature, we adopted a main beam efficiency,
${\eta}_{\rm MB}$, of 0.69 for the JCMT at 230 GHz.  It is shown that the
SMA recovered most of the flux in the velocity range lower than $V_{\rm
LSR}\sim$2 km s$^{-1}$, indicating that most of the blueshifted emission in
this velocity range comes from the compact regions labeled b1 and b2.  The
SMA spectrum is slightly brighter than the JCMT spectrum in several
channels, probably due to the calibration uncertainties in both the SMA and
JCMT data.  In the redshifted velocity range, the SMA recovered $\sim$70\%
of the flux at $V_{\rm LSR} > 10$ km s$^{-1}$.  The flux in the velocity
range close to the systemic velocity is recovered $\sim$60\% except the
channel of the deep absorption dip at $V_{\rm LSR}=$~4.4 km s$^{-1}$.  At
this velocity, absorption caused by the foreground cloud also contributes
to the total absence of emission in the interferometric map because the
JCMT spectrum also shows a deep absorption dip at the same velocity.
%At this velocity channel, absorption caused by the foreground cloud also
%contribute to the  total absence of emission in the interferometric map
%because the spectrum observed with the JCMT spectrum also shows the deep
%absorption dip at the corresponding velocity.

\subsection{CO 3--2}

\begin{figure*}
%\epsscale{.70}
\plotone{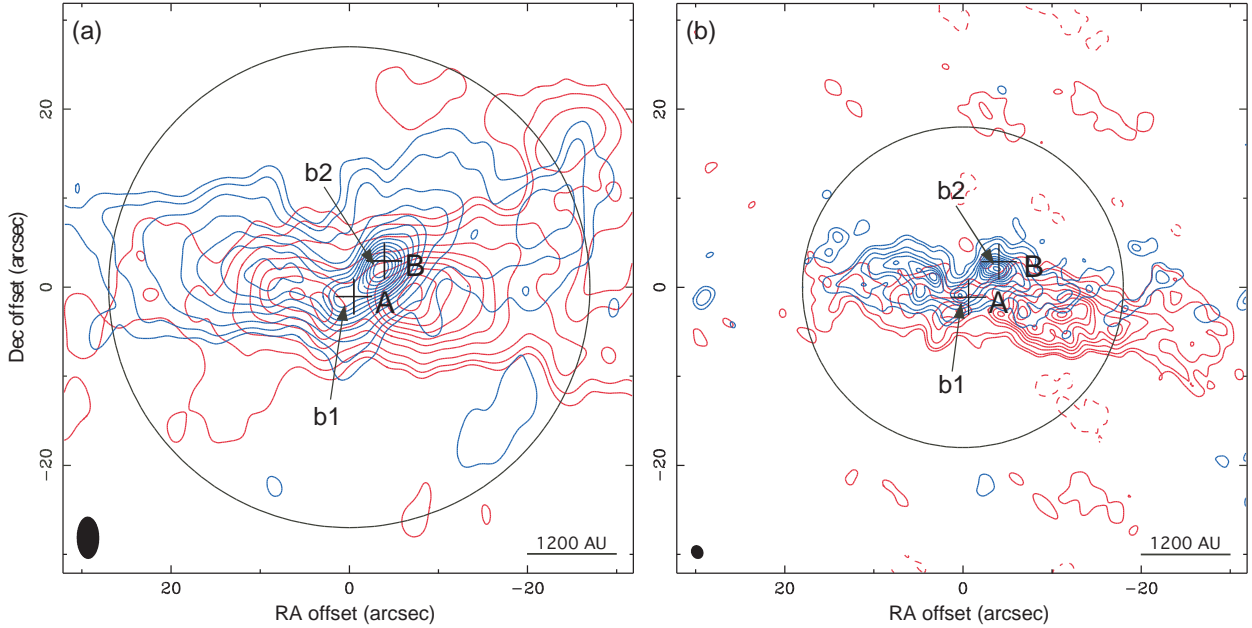}

\caption{{\bf a:} Integrated blueshifted (blue contours) and redshifted (red
contours) CO 2--1 maps.  The emission is integrated over the velocity
channels from $V_{\rm LSR}=-$7.2 km s$^{-1}$ to 3.3 km s$^{-1}$ on the
blueshifted side and from $V_{\rm LSR}=$~5.5 km s$^{-1}$ to 31.9 km
s$^{-1}$ on the redshifted side.  Contours are drawn at $-$5$\sigma$,
5$\sigma$, 10$\sigma$, 20$\sigma$, then every 20$\sigma$ step.  The
1$\sigma$ levels are 0.6 Jy beam$^{-1}$ km s$^{-1}$ for the blueshifted
component and 0.9  Jy beam$^{-1}$ km s$^{-1}$ for the redshifted component.
The two crosses denote the positions of the two continuum sources.
Positions of the two prominent compact components are labeled (b1 and b2).
The open circle indicates the field of view at 230 GHz, and the filled
ellipse indicates the synthesized beam size.  
{\bf b:} Same as Fig. 2a for the CO 3--2.
The emission is integrated over the velocity channels from $V_{\rm
LSR}=-$7.5 km s$^{-1}$ to 3.5 km s$^{-1}$ on the blueshifted side, and from
$V_{\rm LSR}=$5.5 km s$^{-1}$ to 31.5 km s$^{-1}$ on the redshifted side.
Contour levels are 3$\sigma$, 5$\sigma$, and every 5$\sigma$ steps.  The
1$\sigma$ levels are 1.4 Jy beam$^{-1}$ km s$^{-1}$ for the blueshifted
component and 2.3  Jy beam$^{-1}$ km s$^{-1}$ for the redshifted component.
The open circle is the field of view at 345 GHz, and the solid oval
indicates the synthesized beam size.
Note that the primary beam correction is not applied to the data.
} \label{f2}
\end{figure*}

The CO 3--2 total integrated intensity (zeroth moment) map is shown in
Fig.~\ref{f2}b (uncorrected for primary beam attenuation).  The emission is
integrated over the velocity channels from $V_{\rm LSR}=$~$-$7.5 km
s$^{-1}$ to 3.5 km s$^{-1}$ on the blueshifted side, and from $V_{\rm
LSR}=$~5.5 km s$^{-1}$ to 31.5 km s$^{-1}$ on the redshifted side.  There
is no significant emission in the velocity channel centered at $V_{\rm
LSR}=$~4.5 km s$^{-1}$.  The CO 3--2 emission in this velocity range arises
from the extended ambient cloud and is poorly sampled by the uv coverage of
the CO 3--2 data.
%Therefore, we consider that the CO 3--2 emission in this velocity range
%arises from the ambient cloud with spatially extended structure.
There are two bright blueshifted components, one located 2\arcsec~southeast
of source A and the other located 1\arcsec~southeast of source B, that are
likely to be the counterparts of b1 and b2 identified in the CO 2--1 map
(see Fig.~\ref{f2}a).  The peak brightness temperature of CO 3--2 emission
is 39 K at b1 and 55 K at b2.  These two components are connected by faint
emission.  The blueshifted emission extending toward the northeast with a
P.A. of $\sim $70$^\circ$ corresponds to the eastern blueshifted emission
in the CO 2--1 map.  As in the case of CO 2--1, the redshifted emission
extends to the east and west of source A.

\begin{figure*}
\epsscale{.50}
\plotone{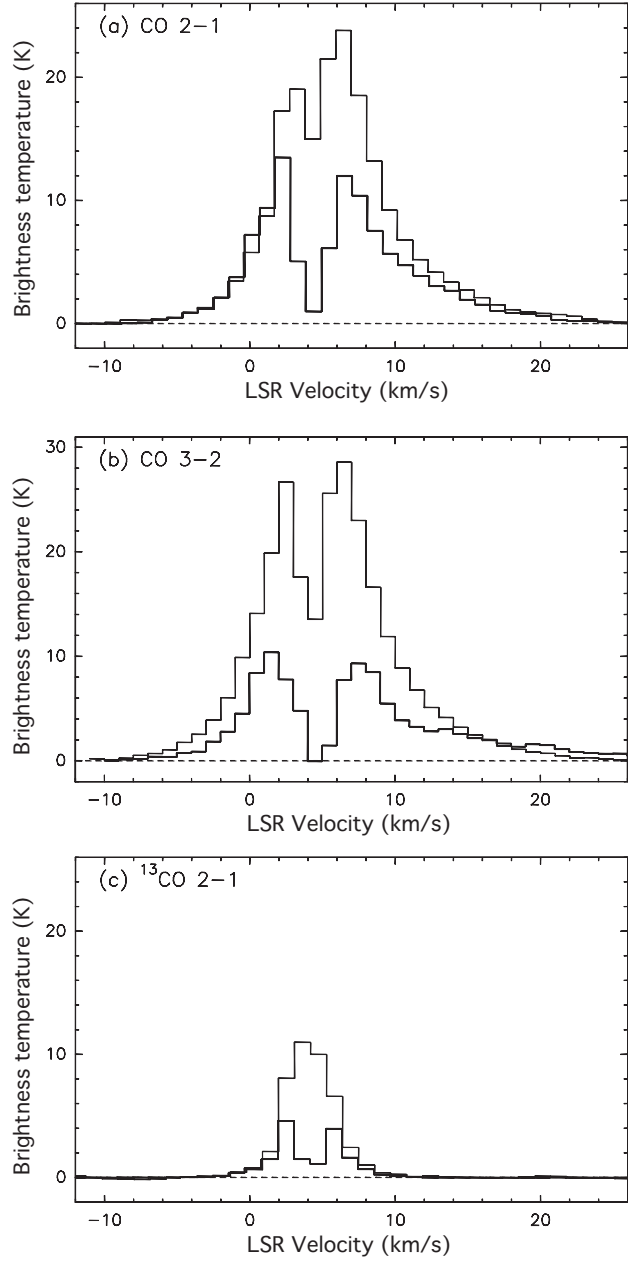}
\caption{A comparison of the CO 2--1 (a), CO 3--2 (b) and $^{13}$CO 2--1
(c) spectra observed with the SMA (black histograms) and the JCMT (grey
histograms) at the pointing center.  The SMA data were first primary beam
corrected, and the spectra were smoothed to have the same angular
resolution as the JCMT data, i.e., 14\arcsec~for the CO 3--2 and
21\arcsec~for the CO and $^{13}$CO 2--1.  The JCMT spectra are resampled to
have the same velocity resolution as that of the SMA data and converted
into brightness temperature by assuming the main beam efficienciies,
${\eta}_{\rm MB}$, to be 0.69 for the CO 2--1 and $^{13}$CO 2--1, and 0.63
for the CO 3--2.  
} \label{f3}
\end{figure*}

In order to estimate the missing flux, we have smoothed the SMA CO 3--2 map
to 14\arcsec~resolution, which is the HPBW of the JCMT at 345 GHz, and
compared the spectrum at the phase tracking center with that obtained by
the JCMT (Fig.~\ref{f3}b).  The primary beam correction was applied to the
SMA CO 3--2 map before it was smoothed.  The JCMT spectrum was converted
into the brightness temperature unit by assuming ${\eta}_{\rm MB}=0.63$ at
345 GHz.  Fig.~\ref{f3}b shows that the missing flux of the SMA CO 3--2
data is larger than that of the CO 2--1 except for the high-velocity part
of the redshifted wing at $V_{\rm LSR}~\ga$~14 km s$^{-1}$, where most of
the flux is recovered by the SMA.  The SMA recovered $\sim$50\% of the CO
3--2 flux in the velocity ranges $-$9.0 $<V_{\rm LSR}<$ 2.0 km s$^{-1}$ and
6.0 $<V_{\rm LSR}<$ 13.0 km s$^{-1}$.  That more missing flux is evident in
the CO 3--2 data than the CO 2--1 data is likely due to it containing fewer
short baselines, as mentioned in \S2.

\subsection{$^{13}$CO 2--1}

\begin{figure*}
\epsscale{1.0}
\plotone{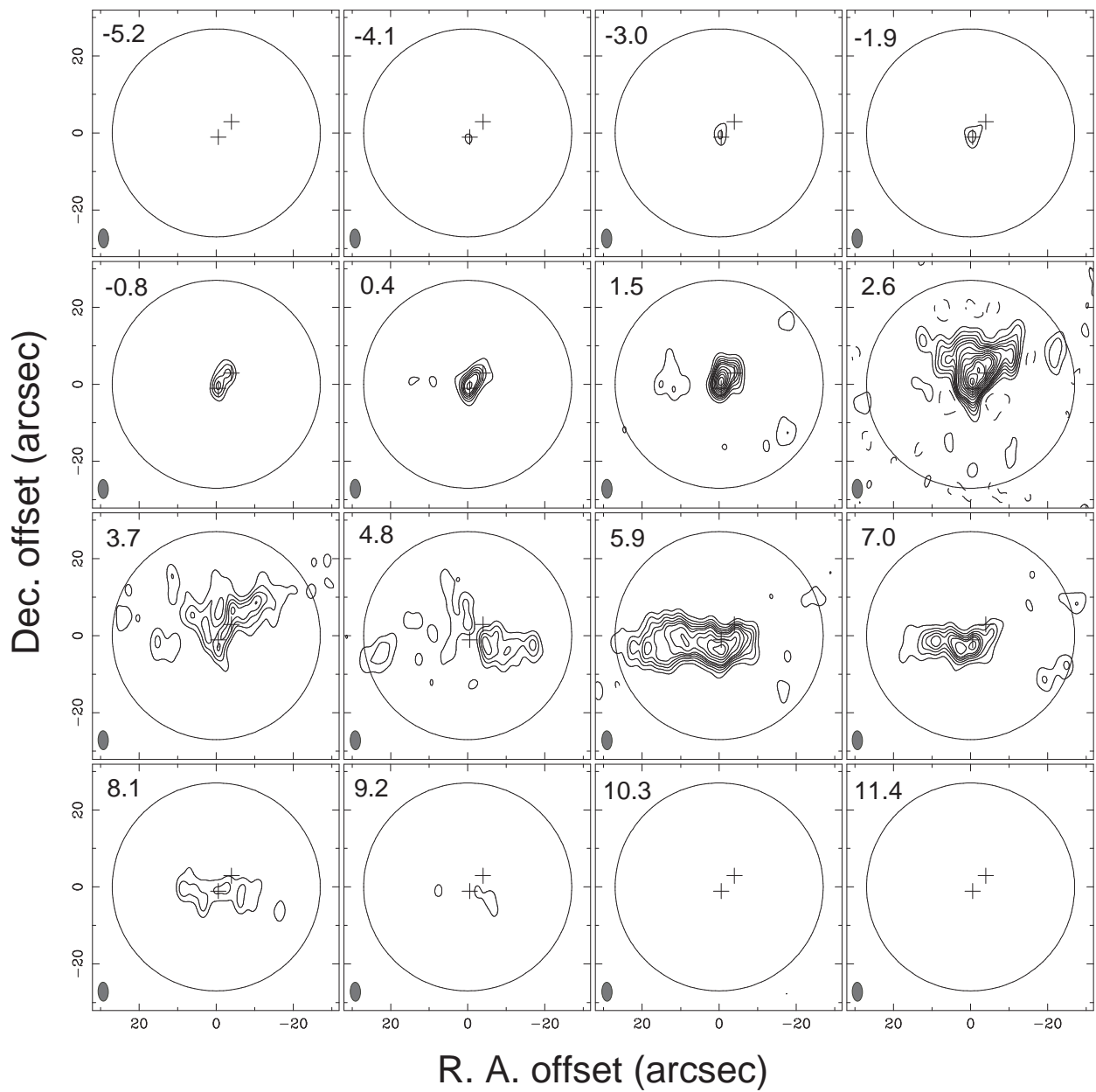}
\caption{Velocity channel maps of the $^{13}$CO 2--1 emission.  The
contours start at $-$5$\sigma$ and are drawn every 5$\sigma$ step until
30$\sigma$, and every 10$\sigma$ in the range above 30$\sigma$.  The
1$\sigma$ level is 0.13 Jy beam$^{-1}$.  The crosses indicate the positions
of the continuum sources A and B; the open circle is the SMA field of view
at 230 GHz. 
Note that the primary beam correction is not applied to the data.
} \label{f4}
\end{figure*}

Fig.~\ref{f3}c compares the $^{13}$CO spectrum observed with the SMA and
that observed with the JCMT.  The SMA spectrum was obtained after we
corrected for the primary beam attenuation and smoothed to
21\arcsec~resolution.  Due to a software problem, the velocity values of
the $^{13}$CO 2--1 data were not recorded correctly.  We found that the
$^{13}$CO 2--1 spectrum observed with the SMA was blueshifted by $\sim$1 km
s$^{-1}$ with respect to that observed with the JCMT.  Therefore, we
calibrated the velocity of each channel in the SMA data using the $^{13}$CO
2--1 spectrum observed with the JCMT.
%Therefore, we referred to the $^{13}$CO 2--1 spectrum observed with the
%JCMT and calibrated the velocity of each channel of the SMA data.
Most of the $^{13}$CO 2--1 flux is recovered by the SMA at $V_{\rm LSR}<$
2.0 km s$^{-1}$.  The velocity channel maps of the $^{13}$CO 2--1 emission
presented in Fig.~\ref{f4} (uncorrected for primary beam attenuation) show
that emission in this velocity range mainly comes from the compact source
close to source A.  It is likely that this component corresponds to b1 in
the CO 2--1 map.  On the other hand, the brightest component b2 in the CO
2--1 map is less significant in the $^{13}$CO 2--1. It is identified as a
northwestern extension of the emission in the velocity channels from $-$0.8
to 1.5 km s$^{-1}$.  The $^{13}$CO emission extends to the east of source A
in the redshifted velocity channels at $V_{\rm LSR}$ = 5.9 km s$^{-1}$ and
7.0 km s$^{-1}$, and extends to the east and west of source A at $V_{\rm
LSR}$ = 8.1 km s$^{-1}$.  It should be noted that a similar structure is
seen in the velocity channel maps of the CO 2--1 (see Fig.~\ref{f1}),
suggesting that the $^{13}$CO 2--1 emission in this velocity range arises
from the dense part of the outflow.  The $^{13}$CO 2--1 flux recovered by
the SMA is 60--70\% in the velocity channels from $V_{\rm LSR}$ = 5.9 km
s$^{-1}$ to 9.2 km s$^{-1}$.  In the velocity channels at $V_{\rm LSR}$ =
2.6 km s$^{-1}$ and 3.7 km s$^{-1}$, the $^{13}$CO 2--1 map shows a
V-shaped structure.  Since the southwestern and southeastern edges of the
$^{13}$CO 2--1 correspond to the northern edges of the western and eastern
lobes, respectively, it is likely that the $^{13}$CO 2--1 emission in this
velocity range comes from the envelope excavated by the outflow.  Note that
the primary beam correction is not applied to the images shown in
Fig.~\ref{f4}. 

\section{Discussion}

\subsection{Outflows in the vicinity of IRAS 16293--2422}

\subsubsection{Driving source}

As was shown in \S3, the CO 2--1 and 3--2 maps observed with the SMA
clearly show small scale outflow emission in the vicinity of I16293. It is
quite natural to assume that the CO 2--1 and 3--2 outflow emission has
the same origin.  Each of the CO 2--1 and 3--2 outflows has a blueshifted
component extending to the east and a redshifted component extending to the
west of source A (see Fig.~\ref{f2}a and \ref{f2}b).  In the CO 2--1
channel maps, both the eastern blueshifted component and the western
redshifted component show fan-shaped patterns with their apexes pointing
toward source A.  Although the eastern redshifted component is less clearly
extending from the position of source A, its coincidence with the eastern
blueshifted emission suggests that the eastern blue- and redshifted
components represent the front side and the back side of a single outflow
lobe, respectively. 
%Although the eastern redshifted component has less clear extension from
%source A position, it positional coincidence with the eastern blueshifted
%one suggests that 
Such an outflow lobe showing both blue- and redshifted emissions together
is likely to be an outflow with a wide opening angle and an axis close to
the plane of the sky \citep{Cabrit86}. These morphological characteristics
of the small scale outflow suggest that it is most likely driven by source
A.

The direction of the outflow axis is rather difficult to define clearly.
The western redshifted lobe in CO 2--1 shows a position angle of almost
$-$90$^{\circ}$, while the blue- and redshifted eastern lobes show position
angles of $\sim$75$^{\circ}$.  The axis of the CO 3--2 outflow is
approximately along a position angle of 80$^{\circ}$.  These results
suggest that the axis of the small scale outflow is approximately the
east-west direction.   The difference in direction between the blue- and
redshifted components of the CO 2--1 emission, and the CO 2--1 and 3--2
emission, could also indicate that the high velocity gas comprises emission
from two or more outflows.  There may be an interaction between the wind or
jet driving the large scale NE--SW outflow and the molecular gas, as
indicated by the offset of shock-tracing molecule line emission peaks to
the NE of source A (Chandler et al.\ 2005).   However, the bulk of the CO
emission we observe is consistent with a predominantly E--W outflow.  

\subsubsection{Comparisons with the large scale outflows}

It is important to know the relationship between the small scale
($\sim$3000 AU) outflow observed with the SMA and the large scale
($\sim$15000 AU) outflows previously observed with single-dish telescopes
even though the size scales are quite different between them.  We should
note that the missing flux and the single field of view in the SMA
observations prevent us from directly relating the two types of outflows on
these very different scales.  Nevertheless, morphological comparisons of
the small and large scale outflows could allow us to infer their
relationship.

Single-dish observations showed two large scale outflows; one is the E--W
pair and the other is the NE--SW pair \citep{walker1988, mizuno1990, Lis02,
stark2004}. The blue- and redshifted lobes of the E--W pair are located to
the east and the west, respectively, showing the same velocity trend as the
small scale outflow.  The outflow axes of these two outflows are also
similar.  These facts naturally suggest that the small scale outflow is the
inner part of the large scale E--W pair, and that the bulk of the CO
emission observed by the SMA is due to the E--W outflow.  If this is the
case, then the E--W pair is not a fossil flow but a currently active one,
and the driving source of the E--W pair is most likely to be source A.

Do the SMA maps show any counterparts of the NE--SW pair?
\citet{chandler2005} proposed that the driving source of the NE--SW pair is
one of the radio sources, A2, based on its bipolar morphology along the
NE--SW direction in their 43 GHz map.  However, our SMA maps show no
counterpart of the NE--SW pair, if we consider the small scale outflow
mapped with the SMA as the inner part of the E--W pair.  One possible
reason for why there is no counterpart of the NE--SW pair in the SMA maps
is that the NE--SW outflow lobes are located outside the SMA field of view
at both 230 and 345 GHz.  In fact, single-dish maps \citep{mizuno1990,
Lis02, stark2004} show that both blue- and redshifted lobes are detached
by $\sim$1\arcmin~from the centroid of I16293.  This implies that the
NE--SW pair is older than the E--W pair, so the NE--SW pair has already
cleaned out the gas along its travel direction.  However this
interpretation is totally different from those of previous studies, where
the NE--SW pair is considered younger than the E--W pair.  On the other
hand, the NE--SW pair is better collimated than the E--W pair.
Observations of CO outflows of YSOs at different evolutionary stages
suggest that the opening angle of the outflow lobe 
%in the vicinity of the central source is small in the earliest phase and 
becomes larger as the central source evolves \citep[e.g.,][]{arc06}.  If we
take this into account, it is natural to consider that the well-collimated
NE--SW pair is younger than the E--W pair.  It is possible that the small
scale counterpart of the NE--SW pair is absent because this outflow is
currently inactive due to its episodic nature.  Eposodic outflows have been
found in several YSOs, such as L1157 \citep[e.g.,][]{gueth96}.
%Episodic nature is known in several outflows such as the one in L1157
%\citep[e.g.,][]{gueth96}.
Since there is no counterpart of the NE--SW pair in our SMA maps, we are
unable to determine if the NE--SW outflow is driven by source A of B, and
to support or refute the different claims by previous studies regarding
this flow \citep{stark2004, chandler2005}.

\subsection{Kinematics of the small scale outflow}\label{model}

Many models have been proposed to explain the relation between molecular
outflows and protostellar jets.  Of these, two scenarios are considered
plausible: (1) the jet-driven model, where molecular outflows consist of
ambient gas swept up by the bow shock of the jet head \citep{raga1993,
masson1992, chernin1994}, and (2) the wind-driven model, where molecular
outflows are the swept-up gas entrained by wide-angle magnetized wind
\citep{shu1991, shu2000, li1996}.  These scenarios have different
kinematical features in a Position--Velocity (P--V) diagram obtained from a
cut along the outflow axis.  The jet-driven model will produce a convex
structure in such a P--V diagram with high velocity components at the jet
head; while the wind-driven model will introduce a  parabolic structure
that originates from the central star \citep[e.g.,][]{lee2000}.

\begin{figure*}
\epsscale{1.0}
\plotone{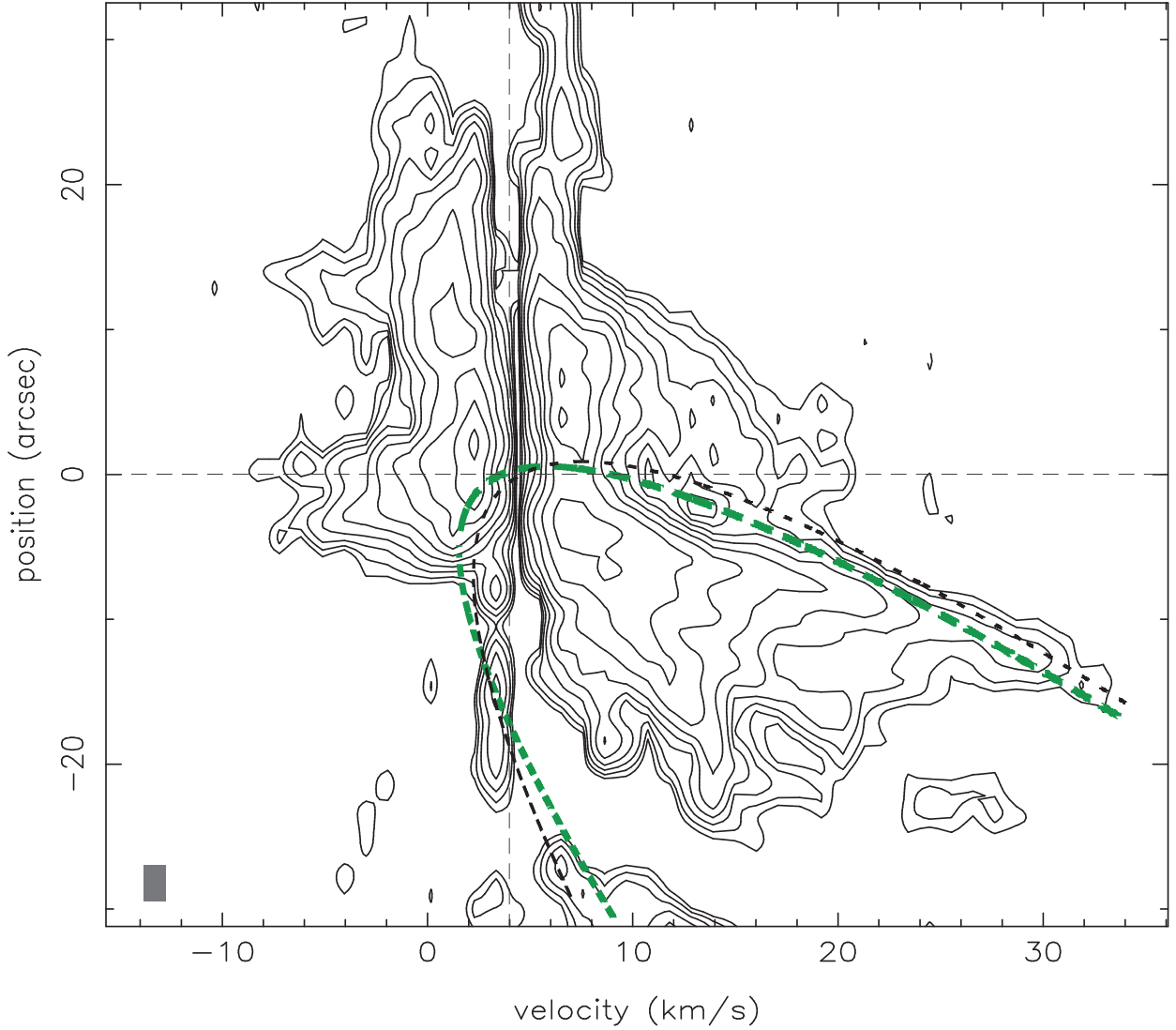}
\caption{Position-velocity (P--V) diagram of the CO 2--1 along a cut at
P.A.~=~90$^\circ$ through the position of source A.  Contours are
logarithmic at: 0.90, 1.43, 2.26, 3.58, 5.68, 9.00, 14.3, 22.6, and 35.8 K.
The horizontal thin dashed line is the position of source A; while the
vertical thin dashed line labels the systemic velocity V$_{\rm sys}$=4.0 km
s$^{-1}$.  The green curve: wide-angle wind model curve produced by {\it
i}=30$^\circ$, $C$=0.20 arcsec$^{-1}$, and {\it v}$_0$=1.30 km s$^{-1}$
arcsec$^{-1}$.  The black curve: wide-angle wind model curve produced by
{\it i}=40$^\circ$, $C$=0.10 arcsec$^{-1}$, and {\it v}$_0$=0.80 km
s$^{-1}$ arcsec$^{-1}$.  The filled square at the bottom left denotes the
spacial and velocity resolution.
} \label{f5}
\end{figure*}

In a CO 2--1 P--V diagram, extracted along the outflow axis at
P.A.~=~90$^\circ$ (Fig.~\ref{f5}), we found that the western redshifted
lobe shows a parabolic shape, as is expected from the wind-driven model.
We therefore adopted  the simplified analytical model of a wind-driven
shell proposed by \citet{lee2000} to study the kinematics of the small
scale outflow.  In the cylindrical coordinate system, the structure and
velocity of the shell can be written as follows:

\begin{equation} 
z=CR^2, {\it v}_R={\it v}_0R, {\it v}_z={\it v}_0z, 
\end{equation} 
where $z$ is the distance along the outflow axis; $R$ is the radial size of
the outflow perpendicular to $z$ ; $C$ and  $v_0$ are free parameters that 
describe the spatial and velocity distributions of the outflow shell,
respectively.

\begin{figure*}
\epsscale{1.0}
\plotone{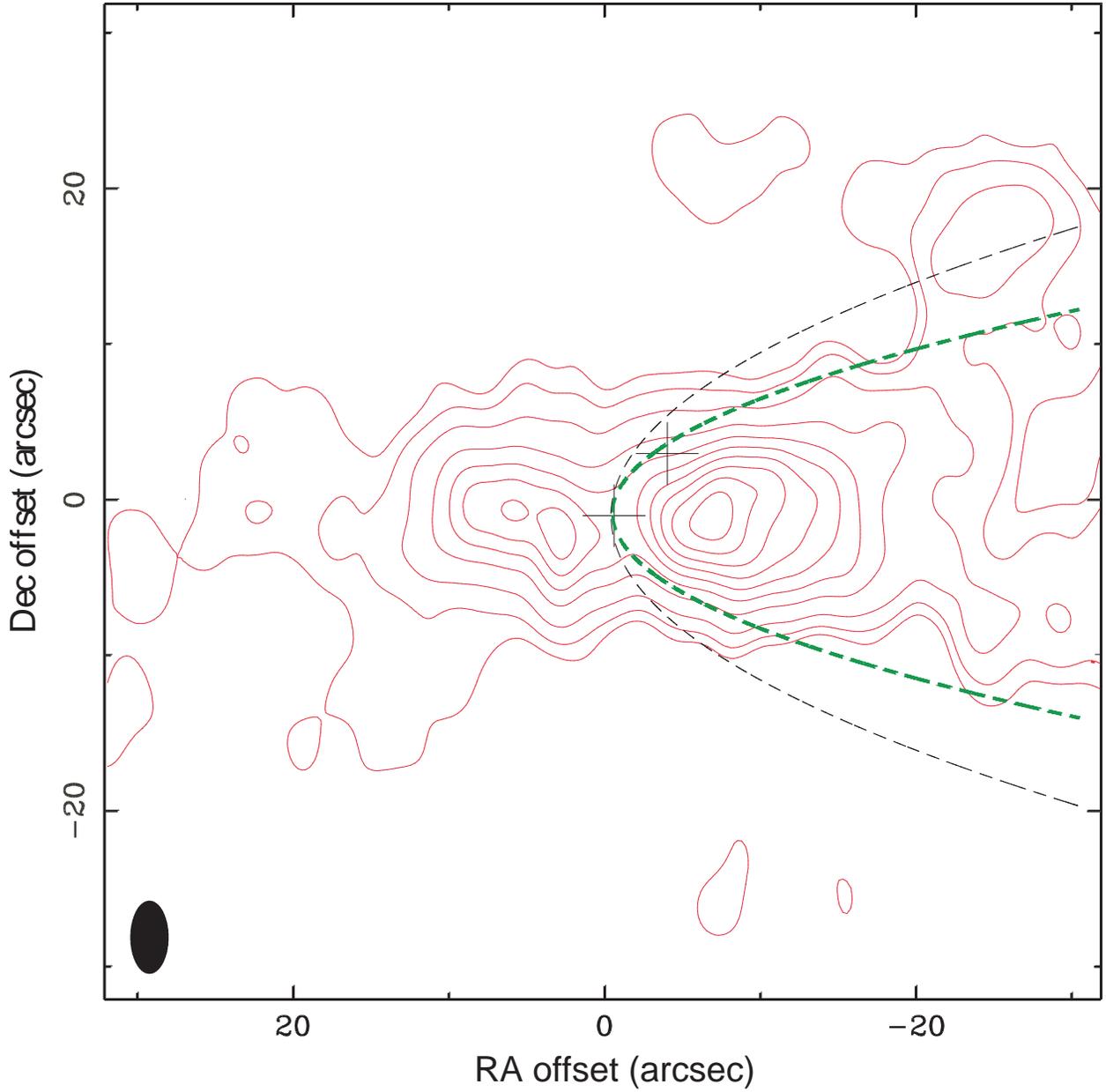}
\caption{Outflow shell model curves overlaid onto the redshifted CO 2--1
emission.  The green line and the black line delineate the curves with
$C$~=~0.20 arcsec$^{-1}$ and $C$~=~0.10 arcsec$^{-1}$, respectively.  Both
curves were produced by assuming the inclination angle of the axis from the
plane of the sky to be 30$^\circ$.
} \label{f6}
\end{figure*}

First, we generated the model curve that delineates the outflow shell
feature\footnote{We only model the red lobe because it shows a clear
fan-shape morphology.} (see Fig.~\ref{f6}), to determine the free parameter
$C$.  In Fig.~\ref{f6} the green line and the black line delineate the
curves produced by $C=$~0.20 arcsec$^{-1}$ and $C=$~0.10 arcsec$^{-1}$,
respectively.  Both curves assume the inclination angle of the axis from
the plane of the sky, {\it i} to be 30$^\circ$.  However, the curve is
insensitive to the inclination angle if {\it i} is smaller than 40$^\circ$.
Since both eastern and western lobes have opening angles of
60$^\circ$--80$^\circ$, and are accompanied by the CO emission component
with opposite velocity shifts (i.e. the eastern blue lobe is accompanied by
the  redshifted component and the western red lobe by the blueshifted
component), it is unlikely that the inclination angle is larger than
40$^\circ$.  Next, we determined {\it v}$_0$ and {\it i} by comparing the
velocity pattern of the model curve with the observed P-V diagram.  The
green curve in Fig.~\ref{f5} was produced by {\it i}~=~30$^\circ$,
$C=$~0.20 arcsec$^{-1}$, and {\it v}$_0$~=~1.3 km s$^{-1}$ arcsec$^{-1}$;
while the dashed curve was produced by {\it i}~=~40$^\circ$, $C=$~0.10
arcsec$^{-1}$, and {\it v}$_0$~=~0.8 km s$^{-1}$ arcsec$^{-1}$.  A
wind-driven outflow model predicts the  ``Hubble law'' velocity structure,
that is the velocity increases as the emission moves away from the central
source \citep[e.g.,][]{shu1991, lee2000}.  As we have mentioned in \S3.1,
the redshifted lobe of the small scale outflow has this signature.  Hence,
our simple modeling qualitatively indicates that the driving mechanism for
the small scale outflow could be a wide-angle wind inclined to the plane of
the sky by $\sim$30$^\circ$ to 40$^\circ$.  The dynamical age of the small
scale outflow is given by 1/{\it v}$_0$ and is 450--700 years.

It should be noted that the wind-driven model does not contradict to the
evidence of a precessing jet presented by \citet{chandler2005}.  Recent
numerical simulations of magnetocentrifugal winds predict an on-axis
density enhancement within the $X$-wind type of wide opening angle wind
\citep{li1996, shang2002, shang2006}, and so it is possible to have a
jet-like feature in such a model that does not show up in CO observations.

\subsection{Physical parameters of the small scale outflow}

We estimated the outflow mass using the CO 2--1 and $^{13}$CO 2--1 data
cubes corrected for the primary beam attenuation and assuming LTE
conditions.  Since the peak brightness temperature of the CO emission
reaches $\sim$50 K, we assumed 50 K as an excitation temperature.  We
adopted an H$_2$ to CO abundance ratio of 10$^4$ and a mean atomic weight
of the gas of 1.36.  A comparison between the channel maps of CO 2--1
(Fig.~\ref{f1}) and those of $^{13}$CO 2--1 (Fig.~\ref{f4}) suggests that
the CO 2--1 emission is significantly optically thick in the low-velocity
redshifted component at  $V_{\rm LSR}\sim$5--9 km s$^{-1}$ and the
low-velocity blueshifted component at $V_{\rm LSR}\sim$0--2 km s$^{-1}$.
The $^{13}$CO 2--1 emission is also detected in the positions of the bright
spots b1 and b2; however, we excluded these components from our
calculation.  The physical conditions of these bright spots are discussed
in the next subsection.  At velocities where the $^{13}$CO emission is
detectable, we estimated the optical depth of the CO 2--1 line using the
following equation:
\begin{equation} 
\frac{S_{\nu}(\rm{CO})}{S_{\nu}(^{13}\rm {CO})} =
\frac{[\rm{CO}]}{[^{13}\rm {CO}]} \left(\frac{1-e^{-\tau}}{\tau}\right),
\end{equation} 
where [CO]/[$^{13}$CO] is the abundance ratio between CO and $^{13}$CO that
is assumed to be 77 \citep{wilson1994}.  The optical depth of the CO 2--1
emission in the low-velocity redshifted component is estimated to be
$\sim$3--36 in the western lobe and $\sim$6--24 in the eastern lobe, and
that of the low-velocity blueshifted component is $\sim$2--17 in the
eastern lobe.  Outflow masses in the optically thick velocity channels were
corrected using:
\begin{equation}
M_{thick}=\sum m(v_i)=m'(v_i)\left(\frac{\tau_i}{1-e^{-\tau(v_i)}}\right),
\end{equation} 
where {\it m$^\prime$}({\it v$_i$}) is the mass at velocity {\it v$_i$}
without optical depth correction; {\it m}({\it $v_i$}) is the mass after
optical depth correction; {\it $\tau_i$} represents the optical depth of CO
2--1 at velocity {\it v$_i$}.  Outside the velocity ranges where the
$^{13}$CO emission was observed, we assumed that the CO emission is
optically thin, and obtained the mass by $M_{thin}=\sum m'(v_i)$.  The
total mass of the outflow was obtained by $M_{total}=M_{thin}+M_{thick}$.
Table~\ref{table1} summarizes the masses of the blueshifted and redshifted
components in the eastern and western lobes.  The optically thick part of
the spectrum dominates the outflow mass calculations.  The outflow masses
are underestimated by an order of magnitude without the correction for
optical depth.  Since significant flux is missed in the low velocity range
due to the lack of short baseline data and absorption by foreground clouds,
the masses derived here are lower limits.  In both eastern and western
lobes, the redshifted component is significantly more massive than the
blueshifted component.  This suggests that dense ambient gas exists behind
the outflow lobes, so that the blueshifted component in the front can
expand rather freely, while the redshifted component in the back needs to
push more material. 

The dynamical parameters of the flow are summarized in Table 2.  Table 2
gives two kinds of values; one is corrected for the inclination effect
($i=$~35$^\circ$) and the other is without this correction (uncorrected).
The inclination angle of 35$^{\circ}$ is chosen based on the kinematical
modeling described in the previous section.  In order to compare the
energies of the small scale outflow with those of the large scale E--W
outflow observed with single-dish telescopes, we estimated the momentum
supply rate and the stellar mass loss rate.  These dynamical quantities can
be derived reasonably well even though the map is incomplete
\citep[e.g.][]{bontemps1996}.  The momentum supply rate and the mass loss
rate of the small scale outflow lobes are estimated to be
$\sim$2$\times$10$^{-4}$ $M_\odot$ km s$^{-1}$ yr$^{-1}$ and
9.0$\times$10$^{-7}$ $M_\odot$ yr$^{-1}$, respectively.  The CO 1--0
single-dish observational result of I16293 \citep{mizuno1990} scaled to a
distance of 120 pc suggests  that the momentum supply rate and the mass
loss rate of the large scale E--W outflow are $\sim$1.6$\times$10$^{-4}$
$M_\odot$ km s$^{-1}$ yr$^{-1}$ and $\sim$8.0$\times$ 10$^{-7}$ $M_\odot$
yr$^{-1}$ in the eastern lobe, while those of the western lobe are
$\sim$1.1$\times$10$^{-4}$ $M_\odot$ km s$^{-1}$ yr$^{-1}$ and
$\sim$5.6$\times$10$^{-7}$ $M_\odot$ yr$^{-1}$, respectively.  Although the
parameters we derived from the SMA data are lower limits, it is likely that
the small scale outflow has comparable momentum and energy with the large
scale E--W outflow.  This may imply that the E--W outflow has been fairly
constant in time, if the compact outflow is the inner part of the large
scale E--W outflow.

\subsection{Nature of the bright, compact components}

\begin{figure*}
\epsscale{0.6}
\plotone{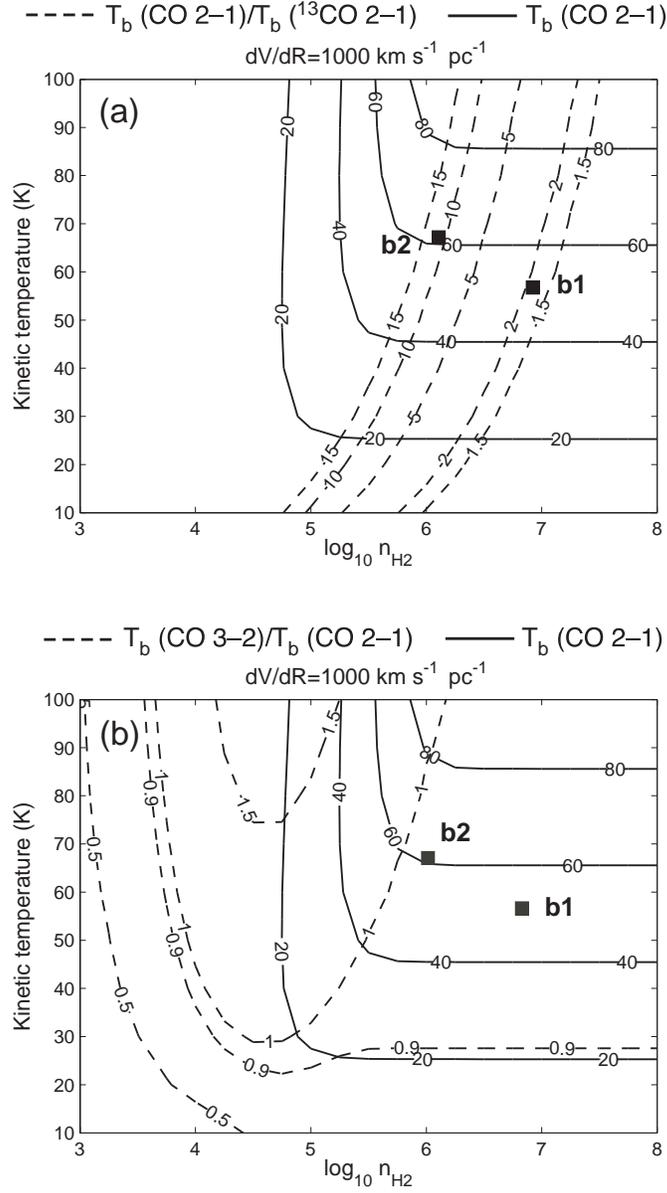}
\caption{{\bf a:} Curves of constant line ratios of CO 2--1/$^{13}$CO 2--1
(dashed lines) and constant CO 2--1 intensity (solid lines) as a function
of the kinetic temperature and the H$_2$ number density for $dv/dr$ = 1000
km s$^{-1}$ pc$^{-1}$.  The filled squares denote the corresponding
observed values of b1 and b2.  {\bf b:} Curves of constant line ratios of CO
3--2/CO 2--1 (dashed lines) and constant CO 2--1 intensity (solid lines).
The locations of b1 and b2 in this diagram correspond to the physical
conditions derived from CO 2--1 and $^{13}$CO 2--1 data.
} \label{f7}
\end{figure*}

As shown in \S3, the high resolution CO maps reveal two bright and compact
components, labeled b1 and b2, at blueshifted velocities (see Fig.~\ref{f1}
and Fig.~\ref{f2}).  In order to investigate the nature of b1 and b2, we
carried out an LVG calculation \citep{goldreich1974, scoville1974} using
the code developed by \citet{oka1998}.  The velocity gradient $dv/dr$ is
assumed to be 1000 km s$^{-1}$ pc$^{-1}$ as estimated from the CO
2--1 line width of $\sim$ 7 km s$^{-1}$ and the size of $\sim$0.005
pc of b1 and b2.  The LVG calculation result is shown in Fig.~\ref{f7}.
Fig.~\ref{f7}a shows the iso-intensity contours of the CO 2--1 line (solid
contours) and iso-line ratio contours of CO 2--1/$^{13}$CO 2--1 (dashed
contours) as a function of kinetic temperature and H$_2$ number density.
Fig.~\ref{f7}b shows the CO 2--1 iso-intensity contours (solid contours)
and iso-line ratio contours of CO 2--1/$^{13}$CO 2--1 (dashed contours).
   
\begin{figure*}
\epsscale{0.6}
\plotone{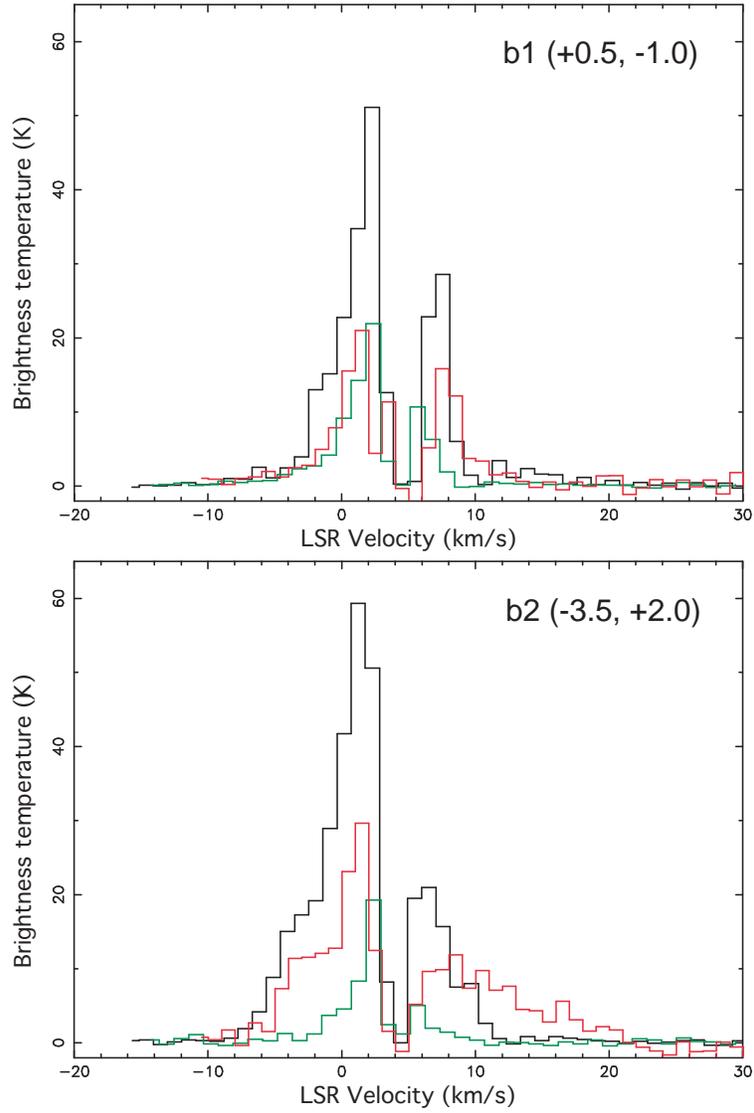}
\caption{Line profiles of the CO 2--1 (black line), $^{13}$CO 2--1 (green
line) and CO 3--2 (red line) at the compact components b1 and b2.  The
position offsets of each component with respect to the phase tracking
center are labeled in arcseconds at the top right. 
} \label{f8}
\end{figure*}

In order to compare the CO 3--2, CO 2--1, and $^{13}$CO 2--1 lines at the
positions b1 and b2, we made the maps of these three lines using the
visibility data in the range from 10 k$\lambda$ to 52 k$\lambda$, 
corresponding to the uv range common for the three data sets.  The beam sizes
in uniform weighting of the CO 3--2, CO 2--1, and $^{13}$CO 2--1 maps were
3\farcs1$\times$2\farcs6 (P.A. 5.2$^{\circ}$), 4\farcs7$\times$2\farcs5
(P.A. $-$3.0$^{\circ}$), and 4\farcs8$\times$2\farcs6 (P.A. 1.5$^{\circ}$),
respectively.  In spite of the same uv range limit, the beam size of the CO
3--2 map is different from those of the CO 2--1 and $^{13}$CO 2--1 due to
differet uv coverages.  Therefore, we convolved the three maps to the same
beam size, i.e. 4\farcs9$\times$2\farcs7 with a P.A. of 0.0$^{\circ}$ after
the correction of the primary beam attenuation.  The CO 3--2, 2--1 and
$^{13}$CO 2--1 spectra obtained at the positions of b1 and b2 are shown in
Fig.~\ref{f8}.  The brightness temperature of CO 2--1 at the position of b1
is $T_{\rm B} \sim$50 K, and that at b2 is $T_{\rm B} \sim$60 K.  The CO
2--1 flux integrated over the velocity range from $V_{\rm LSR}$ = $-$8  km
s$^{-1}$ to 2 km s$^{-1}$ is 102.2 K km s$^{-1}$ at b1 and 207.5 K km
s$^{-1}$ at b2.  The $^{13}$CO 2--1 flux integrated over the same velocity
range is 55.9 K km s$^{-1}$ at b1 and 18.7 K km s$^{-1}$ at b2.  Since most
of the CO and $^{13}$CO 2--1 fluxes in this velocity range were recovered
by the SMA, the effect of missing flux should not be significant.  The CO
2--1/$^{13}$CO 2--1 intensity ratio of 1.8 at b1 suggests that this
component is warm ($T_{\rm kin}\sim$55 K) and extremely dense ($n(\rm
H_2)\sim$10$^7$ cm$^{-3}$).  On the other hand, the CO 2--1/$^{13}$CO 2--1
ratio at the position of b2 is $\sim$11, suggesting that b2 is slightly
warmer ($T_{\rm kin}\sim$65 K) but less denser ($n(\rm H_2)\sim$10$^6$
cm$^{-3}$) than b1.

The CO 3--2 flux integrated over the same velocity range as CO 2--1 is 58.4
K km s$^{-1}$ at b1 and 105.6 K km s$^{-1}$ at b2.  The CO 3--2/2--1 ratio
at the positions of b1 and b2 calculated from the integrated intensity
values are 0.57 and 0.51, respectively.  On the other hand, Fig.~\ref{f7}
shows that the line ratio should be $\gtrsim$0.9 if the CO 2--1 intensity
is brighter than 20 K.  However, in contrast to CO 2--1, only  $\sim$50\%
of the CO 3--2 flux in this velocity range was recovered by the SMA.  If we
assume that most of the CO 3--2 flux within the central 14\arcsec~region
arises from the compact sources b1 and b2, their actual flux values might
be twice the calculated values.  If we take this effect into account, the
CO 3--2/2--1 ratio might be close to 1 at both b1 and b2.  Therefore, the
CO 3--2 results do not contradict the physical conditions inferred for b1
and b2 from the CO and $^{13}$CO 2--1 data.  We should note that there are
two effects that could contribute to a low CO 3--2/2--1 ratio.  If the CO
emission is assumed to be in LTE and optically thick, the CO 3--2 emission
has a higher optical depth than CO 2--1.  Therefore if there are
temperature gradients due to internal heating sources, the CO 3--2 emission
will tend to trace outer and cooler gas.  Even for a uniform temperature,
the $\tau=$~1 surface of the CO 3--2 emission with higher optical depth
will be more extended and more flux will be resolved out by the
interferometer, independent of the other uv coverage issues.

It should also be noted that the kinetic temperature and density derived
here are the averaged values over the beam area of
4\farcs9$\times$2\farcs7.  Since the bright spots b1 and b2 are compact,
the CO 2--1 brightness temperature at b1 and b2 might be higher if we
observe with higher angular resolution.  Indeed, the brightness
temperatures of CO 2--1 and CO 3--2 at b1 and b2 measured in the higher
resolution maps are higher than those of the spectra shown in
Fig.~\ref{f8}.  Therefore it is possible that the kinetic temperatures at
b1 and b2 are higher than those estimated here.
 
Complex molecules, such as HCOOCH$_3$ and C$_2$H$_3$CN, have been detected
toward both source A and source B \citep{kuan2004,chandler2005}.  The
estimated kinetic temperatures ($>$55 K at b1 and $>$65 K at b2) are
consistent with the rotational temperature of $\sim$100 K derived from the
analysis of the SMA HCOOCH$_3$ data \citep{kuan2004}.  
%\citet{kuan2004} and \citet{chandler2005} have detected complex organic
%molecules such as HCOOCH$_3$ and C$_2$H$_3$CN toward both source A and
%source B. 
Since the bright components b1 and b2 are close to source A and B,
respectively, it is likely that these warm and dense components are
associated with hot core-like emission.  Our CO 2--1 and 3--2 maps show
that both b1 and b2 are blueshifted from the systemic velocity and are
offset from the positions of continuum peaks.  These results support the
idea that shocks play an important role in forming hot cores around low
mass protostars \citep[e.g.][]{vanDishoek1995,chandler2005}.  As discussed
in \citet{chandler2005}, it is likely that an interaction between the jet
from source A and extremely dense ($\sim$10$^7$ cm$^{-3}$) gas at b1
contributes to the observed rich chemistry in this region.  On the other
hand, it is unclear how the b2 component is formed.  This component is also
seen in HCN 4--3 emission at blueshifted velocities from $-$2.62 to +0.65
km s$^{-1}$ \citep{takakuwa2007}.  In addition, high resolution H$_2$CO,
SO, and H$_2$S maps presented by \citet{chandler2005} show the local
emission peaks close to b2 rather than the continuum peak of source B.
Since the sulphur-bearing molecules and H$_2$CO are known to be enhanced in
shocks \citep[e.g.][]{bachiller2001}, the component b2 is also likely to be
a shocked region.  If source B is a protostar in the earlier stage than
source A as argued by \citet{wootten1989} and \citet{chandler2005}, b2 may
be a spatially compact outflow from source B that can interact with dense
envelope gas and produce shocks.  Another possibility is that b2 indicates
the shock produced by the interaction between the outflow from source A and
the envelope surrounding source B.

\section{Conclusions}

We have carried out CO 2--1 $^{13}$CO 2--1, and CO 3--2 observations of the
proto-binary system IRAS 16293-2422 with a resolution of
1\farcs5--5\arcsec~using the Submillimeter Array (SMA), which are about a
factor of 5 better than previous observations in these lines.  Our
observations reveal the detailed structures of molecular outflows close to
the binary system.  Our main results are summarized as follows:

1. The high resolution images of CO 2--1 and CO 3--2 show a compact bipolar
outflow along the east-west direction in the vicinity of the binary.  The
center of this small scale outflow is close to source A, suggesting that
source A is  the driving source.  If we interpret this small scale outflow
as the inner part of the large scale E--W pair of the quadrupolar outflow,
the E--W pair is a currently active outflow rather than a fossil flow.  On
the other hand, there is no clear counterpart of the large scale NE--SW
pair in our interferometric maps.

2. The shape and velocity structure of the western redshifted lobe are
well described by an analytical model suggesting that the small scale E--W pair
could be driven by a wide-angle wind with an inclination angle of
$\sim$30$^\circ$--40$^\circ$.

3. There are two compact and bright spots at blueshifted velocities.  One
is located 1\arcsec\ east of source A (labeled b1) and the other is 
1\arcsec\ southeast of source B (labeled b2).  An LVG analysis shows that b1
is extremely dense, n(H$_2$)~$\sim$10$^{7}$ cm$^{-3}$ and warm T$_{\rm kin}
>$~55 K, while component b2 has a higher temperature of T$_{\rm kin}
>$~65 K but slightly lower density of n(H$_2$)~$\sim$10$^{6}$ cm$^{-3}$.
It is likely that these bright spots are produced by means of shocks and
are associated with hot core-like activity observed toward both source A and
source B.

\acknowledgements
We would like to thank all the SMA staff who have helped us to carry out
the observations.  We also wish to thank the anonymous referee who gave the 
insightful comments. 
N. H. is supported in part by NSC grant 96-2112-M-001-023.
Dr.  Y. -N. Su is acknowledged for fruitful discussions
on this research.

\clearpage

\begin{deluxetable}{ccc}
\tablecolumns{3}
\tablewidth{0pc}
\tablecaption{Masses of the small-scale outflow}
\tablehead{
\colhead{} & \colhead{Eastern lobe} & \colhead{Western lobe}\\
\colhead{Component} & \colhead{(10$^{-3}$ $M_{\odot}$)} & \colhead{(10$^{-3}$ $M_{\odot}$)}}
\startdata
Blue\tablenotemark{a}  & 5.4 & 1.1\\
Red\tablenotemark{b}  & 19.1 & 12.5\\
Total & 24.5 & 13.6\\
\enddata
\label{table1}
\tablenotetext{a}{Velocity range: $V_{\rm LSR}$= $-$7.2--3.3 km s$^{-1}$}
\tablenotetext{b}{Velocity range: $V_{\rm LSR}$= 5.5--31.9 km s$^{-1}$}
\end{deluxetable}
\clearpage

\begin{deluxetable}{lccccc}
\tablecolumns{6}
\tablewidth{0pc}
\tablecaption{Dynamical parameters of the small-scale outflow}
\tablehead{
\colhead{}    &  \multicolumn{2}{c}{Eastern lobe} &   \colhead{}   &
\multicolumn{2}{c}{Western lobe} \\
\cline{2-3} \cline{5-6} \\
\colhead{Parameters} & \colhead{{\it i}=35$^\circ$}   & \colhead{uncorrected}    &
\colhead{}    & \colhead{{\it i}=35$^\circ$}   & \colhead{uncorrected}}
\startdata
Momentum (M$_\odot$ km s$^{-1}$) & 10.8$\times$10$^{-2}$ & 6.2$\times$10$^{-2}$ & 
& 9.1 $\times$10$^{-2}$ & 5.2$\times$10$^{-2}$ \\
Kinetic energy (erg) & 12.1$\times$10$^{42}$ &4.0$\times$10$^{42}$ & 
& 21.9$\times$10$^{42}$ & 7.0$\times$10$^{42}$ \\
Dynamical timescale (yr) & 600 & 850 & 
& 500 & 700 \\
Mass loss rate\tablenotemark{b} (M$_\odot$ yr$^{-1}$) & 9.0$\times$10$^{-7}$ & 3.6$\times$10$^{-7}$ & 
& 9.0$\times$10$^{-7}$ & 3.6$\times$10$^{-7}$ \\
Momentum supply rate (M$_\odot$ km s$^{-1}$ yr$^{-1}$) & 1.8$\times$10$^{-4}$ & 7.2$\times$10$^{-5}$ & 
& 1.8$\times$10$^{-4}$ & 7.3$\times$10$^{-5}$ \\
Mechanical luminosity (L$_\odot$) & 0.31 & 7.2$\times$10$^{-2}$ & 
& 0.42 & 9.6$\times$10$^{-2}$ \\
\enddata
\label{table2}
\tablenotetext{b}{The central wind speed is assumed to be 200 km s$^{-1}$. Then we derived the mass loss rate
by assuming momentum conservation of the entrained CO gas and the central wind.}
\end{deluxetable}

\clearpage 

\clearpage 

\clearpage

\clearpage 

\clearpage 

\clearpage 

\enddocument